\documentstyle[12pt]{article}
\def\beq{\begin{equation}}
\def\eeq{\end{equation}}
\def\bea{\begin{eqnarray}}
\def\eea{\end{eqnarray}}
\def\nn{\nonumber}
\def\hp{\hat p}
\def\hd{\hat d}
\def\tp{\tilde P}
\def\td{\tilde D}

\begin{document}
\begin{tabbing}
\` OS-GE-34-93 \\
\` RCNP-057 \\
\` July 1993
\end{tabbing}
\begin{center}
\vfill
{\large Quantum Deformation of the Affine Transformation Algebra}

\vspace{2cm}

N. Aizawa$^*$
\vspace{0.5cm}

{\em Research Center for Nuclear Physics}

{\em Osaka University, Ibaraki, Osaka 567, Japan}

\vspace{0.7cm}
and

\vspace{0.7cm}
H.-T. Sato$^{\dagger}$
\vspace{0.5cm}

{\em Institute of Physics, College of General Education}

{\em Osaka University, Toyonaka, Osaka 560, Japan}

\end{center}

\vfill
\begin{abstract}
  We discuss quantum deformation of the affine transformation algebra.
It is shown that the quantum algebra has a non-cocommutative Hopf
algebra structure, simple realizations and quantum tensor operators.
\end{abstract}
\noindent-------------------------------------------------------------------\\
$^{\dagger}$ {\footnotesize Fellow of the Japan Society for the
Promotion of Science} \\
\mbox{}\hspace{0.3cm}{\footnotesize E-mail~:
hsato@jpnyitp.yukawa.kyoto-u.ac.jp} \\
$^*$ {\footnotesize E-mail~: aizawa@rcnpth.rcnp.osaka-u.ac.jp}
\newpage
%
%
  The notion of quantum deformation of Lie algebras and the groups
has been developed in recent years [1-6]. The quantum groups is
characterized by the fact that the elements of its representation
matrix do not mutually commute. And the quantum algebras is understood
that the universal enveloping algebra has the structure of
non-cocommutative Hopf algebra. The deformation of the classical groups
and the algebras have been discussed by many authors.
The simplest example of the quantum algebra is $ U_q(sl(2)) $ and that of
the quantum group $ SL_q(2) $. Their representation theories have
been well investigated. It is  also recognized that $ U_q(sl(2))$ and
$ SL_q(2) $ are dual each other \cite{tak,maj}.
The deformation of exceptional and super groups and their algebras
are also developed \cite{exc,super}.

  In this paper, we present the quantum deformation of the affine
transformation algebra which is one of the fundamental transformations.
We suggest that the quantum algebra is possible to be provided by deforming
maps from the undeformed algebra.
Some examples of the realizations are also shown.
The quantum analogue of the tensor operator, which carries the adjoint
representation, is
explicitly given in terms of the generators of
the quantum affine transformation algebra.

 The affine transformation means the following operations
\beq
  \hp = -i {d \over dx}, \hspace{1cm}
  \hd = {i \over 2}(x{d \over dx} + {d \over dx} x),     \label{eq1}
\eeq
where $ \hp $ is the momentum operator in one-dimensional quantum mechanics
and $ \hd $ is the dilatation operator.
It is easy to observe that they satisfy the relations
\beq
   [\hp,\,\hd] = i\hp,
\eeq
and
\beq
 \exp{(-ia\hp)}\, x \,\exp{(ia\hp)} = x - a,               \label{eq2}
\eeq
and
\beq
 \exp{(i\ln b\, \hd)}\, x \,\exp{(-i\ln b\, \hd)} = bx.           \label{eq13}
\eeq

 In an abstract sense, the affine transformation is the group whose
multiplication law is given by
\beq
  U(a,b)\, U(\alpha,\beta) = U(a+\alpha b, \beta b).         \label{eq23}
\eeq
The elements of the group can be parametrized by
\beq
  U(a,b) = \exp{(iap)}\,\exp{(-i\ln b\, d)},              \label{eq21}
\eeq
where the Lie algebra $ A = \{p, d\} $ satisfies the commutation relation
\beq
  [d,\, p] = ip.                                          \label{eq22}
\eeq
The allowed region of the parameters is
$ -\infty < a < \infty, \;\; 0 < b < \infty $,
which requires that the
group manifold is a half-plane, and so the affine transformation group is
a non-Abelian and non-compact group.

 The unitary representations of the affine transformation group are found
in refs.\cite{gn} and \cite{ak}. These papers have shown that there are two
and only two unitary
irreducible representations. One is the case of the positive eigenvalues
of $p$ and the other the negative case.
For example, if we take the following realization
of the algebra $A$ on the space of square integrable functions
\beq
  p=k,\hspace{1cm} d=(i/2)(k \partial_k + \partial_k k), \label{eq26}
\eeq
the cases of $ k > 0 $ and $ k < 0 $ give inequivalent irreducible
representations.

 The eigenvalue of $ p $ has continuous value in the unitary irreducible
representation. In order to have a simple matrix representation, we
consider the adjoint representation.
The adjoint representation is not a unitary one because it is a finite
dimensional representation and the affine transformation group is
non-compact. The adjoint representation of the algebra $A$
\beq
   p = \left(
       \begin{array}{cc}
       0 & 0 \\
       -i & 0
       \end{array} \right),\hspace{1.5 cm}
   d = \left(
       \begin{array}{cc}
       0 & 0 \\
       0 & i
       \end{array} \right),                                \label{eq28}
\eeq
gives the adjoint representation of the affine transformation group by
substitution of (\ref{eq28}) into (\ref{eq21})
\beq
   U(a,b) = \left(
       \begin{array}{cc}
       1 & 0 \\
       a & b
       \end{array} \right).                                \label{eq29}
\eeq

  Now, let us present a one-parameter deformation of
the universal enveloping algebra
of $ A $ and the realizations. The quantum
affine transformation algebra $ U_q(A) = \{D,P\} $ is defined by the
commutation relation
\beq
  [D,\, P] = i[P],                                       \label{eq31}
\eeq
where
$ [P] \equiv (q^P - q^{-P}) / (q - q^{-1}) $ and $ q $ is as usual a
deformation parameter. This is a non-cocommutative Hopf algebra as can be seen
as follows. The Hopf algebra mappings, coproduct $ \Delta $, counit $ \epsilon
$
antipode $ S $, are given by
\bea
  & \Delta(P) = P \otimes 1 + 1 \otimes P,        \nn \\
  & \Delta(D) = D \otimes q^{-P} + q^P \otimes D, \nn \\
  & \epsilon\,(P) = \epsilon\,(D) = 0,            \label{eq32}    \\
  & S(P) = - P,\;\; S(D) = -D -i\, (\ln q) [P],   \nn
\eea
and eqs.(\ref{eq32}) certainly satisfy the following axioms of the
Hopf algebra
\bea
  & (id \otimes \Delta) \circ \Delta = (\Delta \otimes id) \circ \Delta,
  \nn \\
  & (id \otimes \epsilon) \circ \Delta = (\epsilon \otimes id) \circ \Delta
    = id,
  \label{eq33} \\
  & m(id \otimes S) \circ \Delta = m(S \otimes id) \circ \Delta = 1 \epsilon,
  \nn
\eea
where $ id $ denotes the identity mapping and $m$ the product of
the two terms in the tensor product ;
$ m(x \otimes y) = xy $. If we define the opposite coproduct $ \Delta'$ by
\beq
   \Delta' = \sigma \circ \Delta, \hspace{1cm}
   \sigma(x \otimes y) = y \otimes x,                   \label{eq34}
\eeq
$ \Delta'(P) $ and $ \Delta'(D) $ also satisfy the same commutation
relation as
(\ref{eq31}). For the opposite coproduct,
$ S(P) $ and the counit are not changed while
$ S(D) $ becomes
\beq
  S(D) = -D + i\,(\ln q) [P].                            \label{eq35}
\eeq

  Next, we show some realizations of the $U_q(A)$. The generators $P$ and $D$
can be formally expressed in terms of the undeformed ones
\beq
  P = p, \hspace{1cm} D = {1 \over 2}({[p] \over p} d + d {[p] \over p}).
                                                         \label{eq36}
\eeq
When the representation of $p$ and $d$ in the Hilbert space is considered,
$p$ and $d$ are hermite operators. The realization (\ref{eq36})
of $ P $ and $ D $ is also chosen to be hermitian
in the same representation space when
$ q $ is real or $ |q| = 1 $. If we require only satisfying the commutation
relation (\ref{eq31}), $ D $ can be simply given by
\beq
  D = {[p] \over p} d.                                        \label{eq37}
\eeq
When the representation and the realization of $A $ have the inverse
$ p^{-1} $ or $ [p] $ are proportional to $ p $, they can be transformed into
those of the $U_q(A)$ by making use of (\ref{eq36}) or (\ref{eq37}).

  Here we show the examples of both cases:

\noindent
(i) The case of having $ p^{-1} $. The realization of eq.(\ref{eq26}) is
transformed into
\beq
  P = k, \hspace{1 cm} D = {i \over 2} ([k]\partial_k + \partial_k [k]),
                                                              \label{eq38}
\eeq
it is easy to verify that the commutation relation (\ref{eq31}) holds.

\noindent
(ii) On the other hand, the adjoint representation of the $U_q(A)$ cannot be
obtained by naive use of the relation (\ref{eq36}) since the
adjoint representation of the undeformed generators (\ref{eq28}) does not have
$ p^{-1} $. However $ [p] $ reduces to be proportional to $ p $, {\em i.e.}
\[
  [p] = \delta p, \hspace{1 cm} \delta \equiv {2 \ln q \over q - q^{-1}},
\]
so we get the adjoint representation of $ P $ and $ D $ with the aid of
eq.(\ref{eq36})
\beq
   P = \left(
       \begin{array}{cc}
       0 & 0 \\
       -i & 0
       \end{array} \right),\hspace{1.5 cm}
   D = \left(
       \begin{array}{cc}
       0 & 0 \\
       0 & i\delta
       \end{array} \right).                                \label{eq39}
\eeq
The adjoint representation of the $U_q(A)$ is the matrices (\ref{eq28})
multiplied
by $q$-dependent factor $ \delta $. The factor $ \delta $ becomes
unity as $ q \rightarrow 1 $.

  Finally, we give the tensor operator
which carries the adjoint representation of the $U_q(A)$.
The definition of tensor
operator of quantum algebras was given by Rittenberg  and Scheunert
\cite{ten} in terms of the representation theory of Hopf algebras.
The tensor operator is generally defined through the following adjoint
action of the Hopf algebra $H$.
The adjoint action of $ c \in H$ on $ t \in H$ is defined by
\beq
  ad(c)\, t = \sum_i c_i\, t\, S(c'_i),                         \label{eq310}
\eeq
where we denote the coproduct of $ c$ by
\[
  \Delta(c) = \sum_i c_i \otimes c'_i.
\]
Writing the $ n \times n $ matrix representation of $ c $ as
$ \rho_{ij}(c) $,
the tensor operators $ \{ T_i, \; i = 1,\,2,\, \cdots,\, n\} $ which
carry the representation $ \rho(c) $ are defined by the relation
\beq
   ad(c)\, T_i = \sum_j \rho_{ji} (c) T_j.                      \label{eq311}
\eeq
Namely, the tensor operators $ \{ T_i \} $ form a representation basis under
the adjoint action. Now in the case of the $U_q(A)$,
we can write down the adjoint action of
$ P $ and $ D $ on $ T \in U_q(A) $
\bea
   & ad(P)\, T = [P,\, T], \nn \\
   & ad(D)\, T = D T q^{P} - q^P T D - i(\ln q) q^P T [P].      \label{eq312}
\eea
We therefore find the tensor operators which carry the adjoint representation
of (\ref{eq39}) :
\bea
   & T_1 = q^{-P} D,   \nn \\
   & T_2 = q^{-P} [P].                                          \label{eq313}
\eea
It is noted that RHS of eq.(\ref{eq312}) reduces to the commutators
and eq.(\ref{eq313}) to $ d $ and $ p $ in the limit of
$ q \rightarrow 1 $.

 In this paper, we showed that the quantum deformation of the
affine transformatio algebra is possible. This fact is non-trivial.
The reason is that the affine transformation has simple but different
structure from the classical Lie algebras. The quantum affine transformation
algebra has the non-cocommutative Hopf algebra structure and its deforming
map and tensor operators were given explicitly.
Since the quantum affine transformation algebra still possesses simple
structure, it would be possible
to use the quantum affine transformation algebra as building block of
other quantum  groups and algebras.
We can
make another algebraic relation which belongs to the present $U_q(A)$, i.e. ,
\beq
   \td \tp - q \tp \td = i \tp,                               \label{eq41}
\eeq
introducing the recombination
\bea
  & & \tp = f(\Lambda) P,               \nn \\
  & & \td = iq^{(\Lambda - 1)/2} [\Lambda]_{1/2},              \label{eq42}
\eea
where $ f(\Lambda) $ is an arbitrary function of $ \Lambda $ provided that
$ f(\Lambda) \rightarrow 1 $ as $ q \rightarrow 1 $ and
\[
   \Lambda \equiv -i {P \over [P]} D,
\]

\beq
   [\Lambda]_{1/2} \equiv {q^{\Lambda /2} - q^{-\Lambda /2} \over
                           q^{1/2} - q^{-1/2}}.
                                                              \label{eq43}
\eeq
The above q-commutation relation (24)is a linear algebra,
which is different from the
original nonlinear algebra (\ref{eq31}). It is thus possible to investigate
linear
representations on the linear basis. This subject should be discussed
in a separate paper.

%
%
%

%
%
\end{document}